# Enhancing the Conventional Information Security Management Maturity Model (ISM[3]) in Resolving Human Factors in Organization Information Sharing

**Oyelami Julius Olusegun**

University Technology Malaysia

Faculty of Computing

Department Information systems

Skudai, Johor Bahru 81310

Jooyelami3@live. utm.my

**Norafida Binti Ithnin**

University Technology Malaysia

Faculty of Computing

Department of Information system

Skudai, Johor Bahru 81310

afida@utm.my

## ABSTRACT

Information sharing in organization has been considered as an important approach in increasing organizational efficiency, performance and decision making. With the present and advances in information and communication technology, sharing information and exchanging of data across organizations has become more feasible in organization. However, information sharing has been a complex task over the years and identifying factors that influence information sharing across organization has becomes crucial and critical. Researchers have taken several methods and approaches to resolve problems in information sharing at all levels without a lasting solution, as sharing is best understood as a practice that reflects behavior, social, economic, legal and technological influences. Due to the limitation of the conventional ISM3 standards to address culture, social, legislation and human behavior, the findings in this paper suggest that, a centralized information structure without human practice, distribution of information and coordination is not effective. This paper reviews the previous information sharing research, outlines the factors affecting information sharing and the different practices needed to improve the management of information security by recommending several combinations of information security and coordination mechanism for reducing uncertainty during sharing of information .This thesis proposes information security management protocol (ISMP) as an enhancement towards ISM3 to resolve the above problems. This protocol provides a means for practitioners to identify key factors involved in successful information sharing. The first one is the identification of all stakeholders to be incorporated into information flow. The second is the integration of the existing information sharing legal frameworks, information sharing protocols, information security standards from the ISO/IEC 27001 and management standard ISO9001 with the existing information security management model (ISM[3]). An experiment was conducted to evaluate the performance of the proposed protocol. The results revealed that interoperability, culture and behavior towards information sharing improved by an average of 10 percent.

**Categories and Subject Descriptors**

**[Information Systems]:** Information Security, Data and Information Sharing
**[Information Security Management]:** Security and Protection

## I. INTRODUCTION

Most recently, the report from the national government for information sharing strategies, (2009), Meyer. (2009) and Rodgers, (2010) has observed moral hazard, poor leadership, inadequate information management practices, a non-sharing culture, the negative behaviors of people towards information as well as confidentiality of information share and the privacy and accountability have been noted as a major factors against information sharing today. In another submissions by Sung Jun Jo and Back-Kyoo Joo (2011) observed that culture, psychological commitment and behavior are antecedents of information sharing, it was noted that the intention of employees to share or not have a role to play in effective information sharing, either negatively or positively. From the strand of information sharing research, researchers above have indicated factors that influence information sharing within an organization and its boundaries and as a result, information and data leakages, information insecurity, lack of compliance and management, loses of confidentiality on information shared have increase tremendously, while interoperability among employee and the system have decreased significantly. To keep up with the recent trend in information security management, organization must build a strengthened and formidable information security management system ISMS for it information sharing Kwon et al. (2007). Information behavior seems to evolve as a result of the interplay between elements in cultural contexts. Information behavior, cultural and behavioral difference of indigenous people proved to be the underlying factor that determines the outcome of information sharing across cultural boundaries. The differences in information behavior of literate and indigenous people can influence the extent to which information is shared across cultural boundaries and can undermine it's successfully accomplishment. Taking into account that cultural contexts as well as information products and services play a significant roles in people's information behavior. It seems obvious that information behavior, organization behavior, culture, confidentiality and trust as becomes a factor to be reckoned with when information sharing is planned across organization.

## II. AIMS AND OBJECTIVES

In this paper, we aimed at developing an information security management protocol (ISMP) as a compliment to the existing and conventional information security management maturity model (ISM$^3$) as an alternative tool that could be helpful to control, understudy human behavior and resolve human factors by building trust among employee's and compliance towards information security policies, rules, laws and regulations for effective information sharing. It also aims at addressing the impact of information sharing failure within an organization and the perceptions organization hold as regard to the management of information security. The question is, (1) How organization would evaluate it current state of information security towards information sharing? (2) What are the factors to consider while creating an information security management towards information sharing? In order to achieve the aim in this paper, we stated the following objectives:

1. To formulate information security management process that could be used to standardized information sharing.
2. To integrate laws and legislation in information sharing, information sharing protocols and relevant ISO27001 and 9001 standard with the existing ISM3 and
3. To develop and propose information security management protocol towards the enhancement of the existing ISM3 as a complement for rendering effective information sharing within organization and its partners.

## III. METHODOLOGY

Our method in achieving this stated objective was divided into phases; the first phase is to acquire data. The data acquisition was done by interview, questionnaire and e-mail correspondence. The central focus in this data acquisition is the departments of human resources, operational and management department. The goal of this first phase is to identify the type of data the organization shared on daily bases.

The second phase is to establishing information security awareness training program (ISAT) as a contributing factor to formulate information security management process that could be use to standardized information sharing as stated in objective 1. The goal of establishing ISAT is to enable us to (1) Indentified all stakeholders to be incorporated into the information flow, (2) design policy and governance for information sharing, (3) develop rules for information and data elements for sharing and(4) determine a common operation system for information sharing as a contributing factors in achieving objective 1.

In the third phase, we intend to evaluate the existing standards, frameworks and legislation. Theses existing standards are: ISO27001 and ISO9001 respectively, while the existing information security frameworks like ISM3 would be consider and finally, the legislation (laws and regulation). This legislation comprises of Data Protection Act (DPA) of 1987, Computer Misuse Act (CMA) of 1990, Privacy Act (PA) of 1985, Human Right Act (HRA) of 1998, Common Law Duty of Confidentiality (CLC), Access to Information Act (AIA) of 1985 and Freedom of Information Act (FOIA) of 2000 and information sharing protocol (ISP) that consist a set of good practices to follow when sharing information. The goal of the third phase is to identify the strength, the impact and the weakness of the existing standards, framework and legislation for the selection purposes and this will serve as an input in achieving objective 2 in this paper.

In the fourth phase, we would integrate and justify the reasons why those components and clauses where selected form the laws and legislation in addition to information sharing protocols and relevant clauses from ISO27001 and ISO9001 standard with the existing ISM3. The selection and integration process would be made easy after indentified the strengths, the impact and weakness of this standards, frameworks and legislation from the third phase. The justification process would also be carefully outlined according to each selected items. The goal of this phase is to identify the right components, clauses and sets of good standards of practices when sharing information is concern.

The fifth phase is the development and enhancement process, in this phase we intend to enhance the conventional information security management maturity model ( ISM3) into information security management protocol (ISMP) as a compliment for rendering effective information sharing within organization and its partners as stated in objective 3. To achieve this, a careful recommendation would be outline based on phase 1 (data source), phase 2 (establishing information security awareness training program), phase 3 (Evaluating the Existing Standards, Frameworks and legislation) and Phase 4 (integration and justification process), the goal of this fifth phase is to enhance the existing and conventional ISM3 from the careful and selected recommendations towards the development and enhancement of ISM$^3$ and to propose information security management protocol, so as to achieve the third objective in this paper.

*A. Case study: the organization of YHLI*

YHLI is one of the leading manufacturing companies with around 2,500 employees in the formal capital of Malaysia (Kuala Lumpur) with over 500 employees in Saber and Sarawak. The company has more than 5 locations around Malaysia. This study focuses on one divisions of the organization on how they share information and what are the challenges they faced as they are trying to create a balance information sharing system. As one of the leading manufacturing company in the south-east of Asia, the organization manufactures chemicals such as the industrial chemicals, food chemicals and agro-allied chemical, paints etc. The company serves a wide range of industries, such as food chemical, pharmaceutical, biotechnology and many more. The company has following different departments: Finance, accounting, marketing, information technology, production, purchase, customer service and the human resources department that take cares of employees data. Such data involve personal, non-personal, sensitive and non-sensitive data. The organization shared information with external party such as the Malaysian Health Department that provides medical services to the organization, Insurance firm that insured the legal property and its employees, banks (financial institution) that relate to the employees loans etc and the stakeholders. The stakeholders are people who have direct or indirect interest and shares within the organization and they need to share information with the organization regarding there annual dividend and other capital or investment benefits.

In addition, the information technology (IT) Department of YHLI Company has more than 20 employees serving around 100 users. Apart from employees, there are 4 consultants working on SAP implementation. The company has successfully implemented SAP in 2008 for all major business functions and is currently using it as their ERP system.

This organization was chosen as a subject of analysis due to its geographical location and the enrichment in terms of multi-cultural and devise ethics group and to serve as the specific in-depth case study for the investigation on how human factors that could influence the sharing of information and what factors to be consider when planning and implementing information security management for effective information sharing and data exchange across the organization.

## IV. DATA ANALYSIS

In this session, there is a need to analyzed and interpret the dataset from the data entry. The data entry was done on a Microsoft Excel after correlating the entire questions attempted from all respondent. It should be noted that, the questionnaires were distributed at random to the employees of YHLI. From the survey identity, the total numbers of respondent is 35 and there are 37 questions in the questionnaire, question 1 to 25 focuses on culture and behavioral questions, information security questions, education, training and awareness questions, experimental validation questions respectively while, questions 26 to 37 is an interview question. In details, question 1-8 based on the culture and behavior of employees. This question is tag dataset A, table 1 below shows the frequency table of responses to the questions as dataset A.

**Table 1:** Culture and behavioral questions (dataset A)

| Response | Frequency | | | | | | | | |
|---|---|---|---|---|---|---|---|---|---|
| No | Q1 | Q2 | Q3 | Q4 | Q5 | Q6 | Q7 | Q8 | Total |
| Yes | 32 | 12 | 16 | 10 | 22 | 17 | 11 | 10 | 130 |
| No | 3 | 14 | 10 | 16 | 9 | 5 | 15 | 5 | 77 |
| Blank | 0 | 9 | 9 | 9 | 4 | 13 | 9 | 20 | 69 |

Key: Q means question

From table 1, out of the total number of the questionnaire, this frequency table tag (dataset A) comprises only questions (Q) 1 to 8. It should also be noted that, the blank dataset will be considered as a missing values; hence, it will not be relevant in this thesis. Question 1(Q1) one indicate there believe in culture and out of 35 responses to question one, 32 says yes, 3 says no. This indicates that almost all employees believe in culture as a way of life. Question two is know if culture and there individual believes influence there behavior towards others, in response, 12 says no, while 14 says yes, indicating that the particular respondent to Q2 do not influence others with his/her culture and behavior , the other 12 influenced other with there culture and behavior . Observing Q3, this is to figure out if the employees share information among same culture, the dataset A indicates that, 16 responded with yes while 10 says no which means that some employee shared information based on same culture and believes, it also means that, they are more open-up to other employees with same culture. In Q4 the responses are 10 for yes and 16 for no, this also indicate that, most of the employees are easily going among other cultures, believes and ethnic groups while the other 10 are not. In Q5, Twenty-two (22) have been influenced by other culture and believes while 9 responses negatively, this means that, they are not influence by other culture. Looking at Q6 subset of Q5, 17 responded that the impact affect there sharing behaviors towards others while 5 says no. In Q7, eleven (11) says they have been able to influence others with their own culture, while 15 says No, and in question 8 (Q8) 10 responded that those influence by them developed negative information sharing behavior toward others and 5 remain negative towards the question. Figure 1 show the bar chat that further illustrate dataset A.

From figure 1, the total response to yes in dataset A is 130. In dataset A analysis, yes means that, the organization lack culture and behavior towards information sharing and also there is a general indication regarding culture and behavior show that, culture and behavior plays important roles in information sharing. The organization must initiate a common ground for culture and behavior among employees as it could strengthen information sharing therefore, there is a need to improve on culture and the behavior of employees. To summarize dataset A, the word yes means, the organization need to improve on culture and behavior towards information sharing as the organization is a multicultural environment.

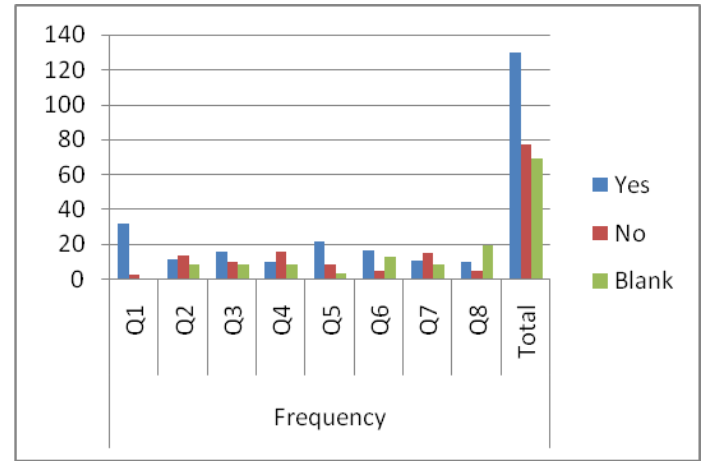

Figure 1: Bar chart of dataset A

## V    EXPERIMENTAL VALIDATION

Under this session, there are needs to further explain experimental validation and its significances in this research. Experimental validation could be the extent of which the finding is genuine and due to the independent variable been manipulated, sometimes, it is hard to interprets the information that where obtained from the interviews and questionnaire due to social desirability bias, this is the tendency to provide socially desirable rather than honest answers during the interviews and on the questionnaires another challenges in experimental validation is the complex interactional process and self-fulfilling prophecy. This self-fulfillment prophecy might be seen as a tendency for someone expectation about another person to lead to the fulfillment of those expectation, that means depending on another person to fulfill a task. These three factors may have influence on our data collection, to avoid this, we perform an experimental validation to determine the genuineness of the data we collected. Table 1.2 shows the frequency of the response of the employees who attempted the questionnaires, this dataset will be tagged datasets D. It is to note that 35 employees responded to this experimental validation questions. Table 2 below shows the frequency in the responses to the experimental validation question.

**Table 2**: Experimental Validation questions (dataset D)

| Response | Frequency | | | | | |
|----------|-----------|-----|-----|-----|-----|-------|
| No of Qs | Q21 | Q22 | Q23 | Q24 | Q25 | Total |
| Yes | 5 | 1 | 28 | 29 | 5 | 68 |
| No | 25 | 27 | 0 | 0 | 23 | 75 |
| Blank | 5 | 7 | 7 | 6 | 7 | 32 |

From the table 2, the experimental validation question carries only 5 questions, Q21 to Q25; these questions are structured towards the sincerity and confidence of employees who contributed in answering the questionnaire. This would enable the researcher to study the variables and to judge the experiment has been done without prejudice, fear and social desirability bias. This frequency table for this dataset D revealed that, only 5 employees responded positively to Q21, 25 employees say No and 5 is blank (missing values). In Q22, 27 says No, 7 blank, while only 1 employees responded positively meaning that, only 1 employee would not entertain fear even when been observed. In Q23, almost all the employees responded positively with 28 showing that, they are honest enough during the filing of the questionnaire, while in Q24 employees yes frequency 29 , this also indicate that, they have answer the questions to the best of there knowledge. Observing the last question in the table 2, 23 employees where negative with response of 23 revealing that, they are not afraid that there commitment towards transparency might affect there job while 5 say yes and 7 blank. From this dataset, it will be concluded that, employees where honest enough during this data collection and the answers given to the best of there knowledge. The bar chart in figure 2 will further illustrates on the overall result of dataset D.

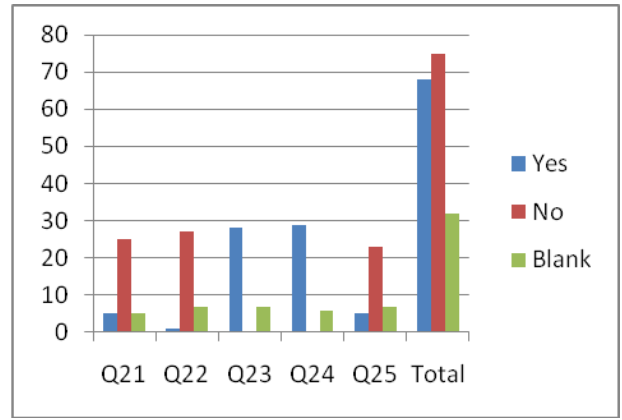

**Figure 2**: Bar Chart for Datasets D

Observing this figure 2, No is the highest, this shows that an average employees responded without been influenced by the demand characteristic of the situation, even they have been observed, it does not influence there behavior towards answering the questionnaire, they have been honest enough during the filing of the questionnaire, they have been able to answer the questions to the best of there knowledge and they where not afraid that the commitment made towards transparency might affect there job.

## VI.  FINDINGS

It is observed from the data analysis that, information sharing does not take place adequately due to lack of leadership, information management practices that restrict sharing of information, privacy and accountability concern, absence of clear value proposition, culture that probably resistant to sharing information and inadequate security education, training and awareness and as a result, this have brought negative impart to the organization in respect to   loses of interoperability, the ability of systems and employees working together if there is no good culture and behavior towards information sharing, loses of confidentiality, information leakages, information insecurity, financial and data loses would be inevitable to the organization information assets.

## VII. IMPLEMENTATION

In achieving the first objectives, what we did was to first evaluate the current information security practices of the organization and we compared and contract it with the data we collected for analysis, from the analysis, we discover that, their are lapses in the organization information security system, although data analysis revealed that, the organization have some considerable information security policies and framework but could not resolve the human error that where observed. This Human error is contributed by negative culture and behaviors towards information sharing, lack of trust, confidentiality, improper management responsibility towards information sharing, privacy and accountability, lack of leadership roles in information sharing coordination etc. that might result into fraud (Illegal alteration of  information for selfish interest) and corruption (wilfully revealing information to unauthorized person for selfish interest) and Incompetence as a result of insufficient information and security education. We then indentify all the stakeholders to be incorporated into information flow of the organization, develop rules and data elements for sharing and the exchange of information to determine a common operating system. The identified stakeholders are: the banking and insurance institution, health ministry, education ministry, external distributors and the customers. With this, we are able to establishing a common trust by improving (1) the employee's act of collaboration and (2) build trust and interoperability.

In achieving the second objective, what we did is to integrate the carefully selected clauses from the ISO27001 and 9001 standards, legal framework and a set of good practices from the information sharing protocol to enhance the conventional ISM3 model to address the human error in information sharing. Then we developed a culture that will rewards information sharing behaviours. This was achieved by promoting mechanisms for sharing information; This mechanism requires the development and execution of the information sharing strategic implementation plan (ISSIP). This implementation planning and execution will occur at all levels. The success of sharing will be a unified and coordinated set of initiatives will span from department leadership to system owners, operators and other entities that share information with the organization. With this, we are able to remove obstacles and welcome better tools that help the organization to succeed in sharing information.

The third objective was achieved with a careful recommendation. These recommendations were noted during the process in achieving the stated objective. Those recommended clauses; standards and frameworks are used in the enhancement process. It is believe that, the recommendations will guide the organization towards effective information sharing within the organization and its external partners. It should also be noted that the recommendations were also induced after careful studies and analysis of the datasets, close-ended interview and the existing information security of the organization. With the recommendations, we are able to propose an information security management protocol (ISMP).

## A. The Proposed Framework

This proposed information security management protocols (ISMP) consist of the selected information security standards from the ISO27001 and quality management system from the ISO9001, these are the two basic standards for information security and effective management respectively while the information agreement and protocol are sets of procedures to follow when considering sharing information within and across the organization. Looking at information sharing legal frameworks served as a legal requirements when sharing information that relate to personal and sensitive data, this also serve in protecting the interest of those who involves in the sharing of information while culture and trust will assist the organization in the establishment of a common trust among employees and developed culture that will reward information sharing within the organization. Indemnity serves as a measures towards punishment to any party that breach the agreement made before and after sharing information, while consent is a kind of documents sign and agreed to by both party who intend to share information or who involves in the sharing of information, in this regards policies and governance in the sharing of information will be clearly stated and the identification and development of rules and data element for sharing will also spelled out clearly. The confidentiality agreement and statement is to indicate to any party sharing the information that, the information been shared is accurate and there is no loss in value of those information. Education, training and awareness is to set the pace for information and security education, this enable the employees to understand why some information are sensitive and also to distinguished all data elements from one another, it will also enable employees to handle information accurately.

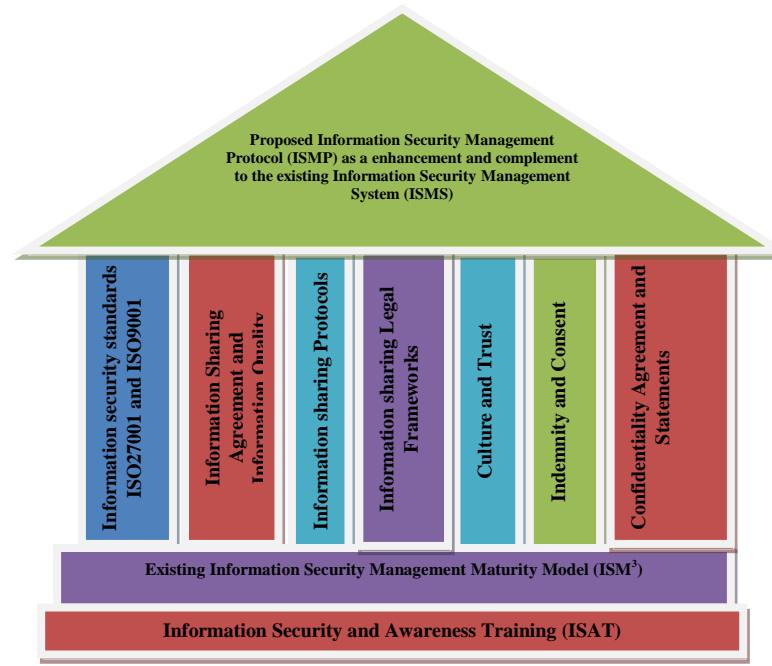

**Figure 3:** The propose ISMP

## VIII. FUTURE WORK

The first step in this direction would be a survey to all certificate users of the ISM3 framework focusing on the use of the ISM3 tool, the perceived efficiency, and reliability etc, a study like this, in cooperation with IT organization in its designing phase and interview with information sharing experts will stir up a new research area. Another idea that has grown stronger during this study is to interview acknowledged experts on the management of information security in organizations, and to analyze these interviews in line with the ideas of grounded theory of information sharing, communication and technology (ISCT) to search for themes and patterns in their views on the issue at hand. In the case study of organization, most of the materials presented in this paper are general, in the sense that, it does not directly tackle the chaotic reality in which organizations have to try to resolve their information sharing problems. In the practical situation, therefore, it would be very valuable to study various organizations in their efforts to manage information security towards information sharing.

## IX. CONCLUSION

Further, this paper reviews the historical and academic contribution on information sharing over the last 30 years. We started with the review of articles in information sharing and the attempt and effort to define information sharing as related to organization and some existing models, frameworks and standards that are related to information security management in the quest to develop an enhancement protocol for information sharing. This paper also highlights the various legislations that have been passed into law to moderate information during sharing and after.

It introduced the concept of information security management protocol (ISMP) as a techniques based on the existing information security model, standards and information sharing legislation (the laws that governs information in the sharing context. Some information sharing expert and researchers like Hepworth, (2007), Williams et al. (2008), Sung Jun Jo and Back-Kyoo Joo (2011), Gary Rodgers, (2010) and many others have seen the need to secure and improve information sharing in organization through positive behaviour, good culture and management practices, while some other researchers like Constant et al. (1994), Brown and Duguid, (2000), Fulk et al., (1995) believe that information sharing is base on social, norms, believe , culture and behaviours, it is also noted from the trend of literature that challenges confronting information sharing is basically on human and researchers like Albert and Barabasi, (2002), Newman, (2003), Kelle and Abrials, (2007), Razavi, (2006), Rafaeli and Raban, (2005). indicated that, information sharing challenges is not to be solely addressed by technology but through a non technical aspect based on human practices, sharing behaviour, culture that encourage sharing of information with the help of leadership and compliance.

ABOUT THE AUTHORS

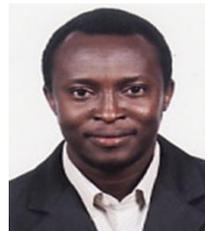

Oyelami Julius Olusegun Is a graduate of electrical engineering technology with electrical power option from college of science and technology Ghana in collaboration with French institute of technology in 1996. Having served in engineering industry for over 10 years, in the quest for computer knowledge, he enrolled into kursk state technical university, Russia where he obtained a B.Sc. in computer system and network engineering in 2009 and several professional certificates in IT with over ten years industrial experience in engineering. Currently is a postgraduate research student in department of information system, faculty of computing, University Technology Malaysia (UTM), and a member of information assurance and security research group (IASRG-UTM), His research interest are in information security management, social networking and information sharing and Information System. He is a professional member, association for computing machinery (ACM) and an academic member, association for information systems (AIS). He has recently extended his research interest into ICT, cloud and grid computing.

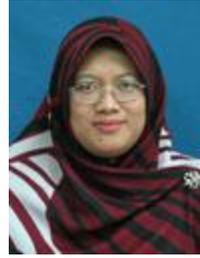 Norafida Binti Ithnin is currently a senior lecturer and head of department in Universiti Teknologi Malaysia (UTM), faculty of computing. She received her B.Sc degree in Computer Science (Computer Systems) form Universiti Teknologi Malaysia (UTM), Kuala Lumpur, Malaysia in 1995 and her M.Sc degree in Information Technology (Computer Science) from Universiti Kebangsaan Malaysia (UKM), Bangi, Malaysia in 1998. She bagged her PhD degree in Computation from UMIST, Manchester, United Kingdom in 2004. Currently, her main research interests are security management and graphical password.